## Title

Quantum Sensing of Weak Radio-Frequency Signals by Pulsed Mollow Absorption Spectroscopy

## Authors

T. Joas, A. M. Waeber, G. Braunbeck and F. Reinhard*

Technische Universität München, Walter Schottky Institut and Physik-Department.

*friedemann.reinhard@wsi.tum.de

## Abstract

Quantum sensors, qubits sensitive to external fields, have become powerful detectors for various small acoustic and electromagnetic fields. A major key to their success have been dynamical decoupling protocols which enhance sensitivity to weak oscillating (AC) signals[1]. Currently, those methods are limited to signal frequencies below a few MHz.

Here we harness a quantum-optical effect, the Mollow triplet splitting of a strongly driven two-level system, to overcome this limitation. We microscopically understand this effect as a pulsed dynamical decoupling protocol and find that it enables sensitive detection of fields close to the driven transition. Employing a nitrogen-vacancy center, we detect GHz microwave fields with a signal strength (Rabi frequency) below the current detection limit, which is set by the center's spectral linewidth $1/T_2^*$.

Pushing detection sensitivity to the much lower $1/T_2$ limit, this scheme could enable various applications, most prominently coherent coupling to single phonons and microwave photons.

## Text

Sensitive detectors for weak radio-frequency (> 100 MHz) signals of electric, magnetic or pressure fields would shift several frontiers of physics. They could advance the exploration of phonons on the single-particle level and reveal weak microwave signals encountered in quantum information processing, biomedical imaging or, more exotically, the search for extraterrestrial intelligence[2].

Driven by this perspective, the past decade has seen the rise of detectors based on Rydberg atoms[3,4], superconducting quantum circuits[5–8] or optomechanical sensors[9–11]. All approaches have achieved noise levels below 10 photons (noise temperatures below 100 mK), an order of magnitude better than state-of-the-art semiconductor detectors[12] and maser amplifiers[13–16]. However, this performance is only reached in sophisticated setups (Rydberg atoms) or at sub-Kelvin temperatures (optomechanics, superconductors).

Detectors based on solid state spin qubits could potentially overcome these limitations. Optically active spin qubits such as nitrogen-vacancy (NV) centers can be optically polarized, that is effectively laser cooled to a temperature of a few 10 mK, even in a substrate at higher temperature. Magnetic tuning of their spin transition enables resonant coupling to external fields at any frequency up to 100 GHz. Theory proposals (Fig. 1a) suggest that both single microwave phonons[17] and photons[18] can be coupled sufficiently strongly to drive a full spin flip within the spin coherence time $T_2$ (ms[19] to s[20], depending on species and temperature).

However, radio-frequency sensing by spin qubits is currently precluded by a major roadblock. It is illustrated in the detection protocol of Fig. 1b, where an incoming signal drives the qubit transition, inducing a spin flip which is subsequently detected by readout of the spin. To drive a full spin flip, an incoming signal has to saturate the spin transition. Therefore, the signal strength (Rabi frequency) has to exceed the inhomogeneous transition linewidth $\Delta\omega \sim 1/T_2^*$. Since $1/T_2^*$ is much broader than $1/T_2$ (MHz vs kHz for an NV center in a natural abundance crystal at room temperature), single spins currently cannot detect the weak signals mentioned above.

For signal frequencies below a few MHz, dynamical decoupling protocols can break this limit (Fig. 1c). Here, the transition is driven by a strong continuous or pulsed control field (frequency $\omega_0$) to create a pair of photon-dressed qubit states, split by the driving field Rabi frequency $\Omega$. This new transition has a far narrower linewidth $1/T_2$, and can hence absorb much weaker signals. However, practical limitations on drive power limit the frequency range which can be probed in this way.

As the key idea of this work, we note that the fully hybridized spin-photon states (the 'Jaynes-Cummings ladder') support another set of transitions at frequencies ($\omega_0 - \Omega, \omega_0, \omega_0 + \Omega$), the Mollow triplet[21], which has been extensively studied in quantum optics[22–26] and has been proposed as a narrowband tunable photon source[27]. Since these transitions equally link pairs of dressed states, we posit that they should allow for $T_2$-limited sensing of radio-frequency fields.

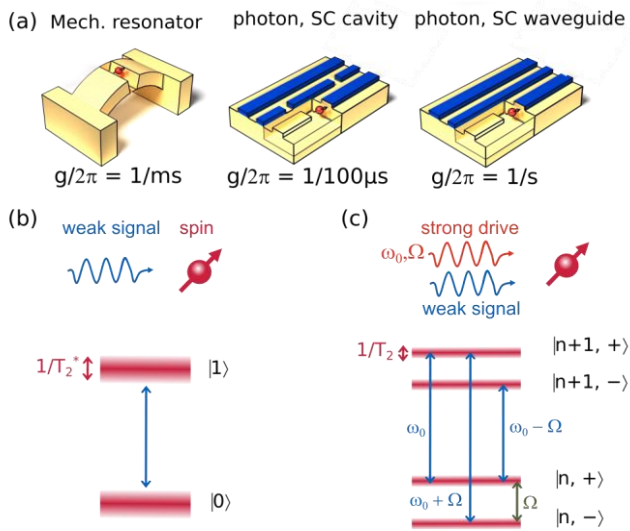

Fig. 1 – High-frequency sensing by Mollow resonance spectroscopy. (a) Examples of signals which could be coupled to spins within $T_2$, but which are out of reach of current $T_2^*$-limited protocols[17,18] (g: coupling strength, SC: superconductor). (b) High-frequency signals driving the bare spin transition have to be stronger than the inhomogeneous linewidth $1/T_2^*$. (c) Dressing of the spin by a strong drive creates a set of narrow transitions at frequencies $\omega_0, \omega_0 \pm \Omega$, which enable $T_2$-limited sensing.

We turn this idea into a sensing protocol by the scheme of Fig. 2a. Here, the spin is initialized into the dressed state $|+\rangle = (|0\rangle + |1\rangle)/\sqrt{2}$ by a $(\pi/2)_Y$ pulse (Y labeling the carrier phase $\phi_Y = \pi/2$). This state is locked as an eigenstate of a strong dressing field with orthogonal carrier phase $\phi_X = 0$. We find that a weak signal field at the detuned frequency $\omega_0 \pm \Omega$ indeed induces rotation at its native Rabi frequency, as evidenced by measurements on an NV center. The central Mollow resonance is absent in these measurements, since it couples dressed states with identical spin projection, as has been previously observed in superconducting qubits[28]. It can be recovered by preparing into an orthogonal state and changing the phase of the signal (Fig. 2b) to account for the different quadratures of the Mollow sidebands[22].

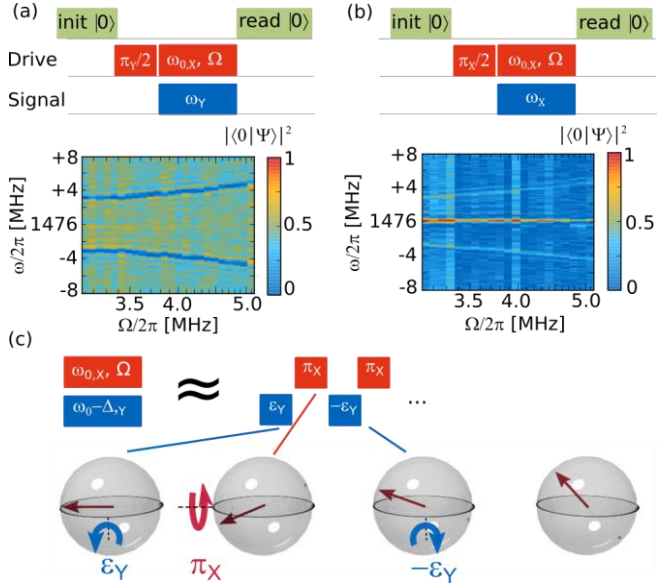

*Fig. 2 – Mollow absorption as a sensing protocol. (a) A weak signal with phase orthogonal to a strong drive field drives the sideband transition at $\omega = \omega_0 \pm \Omega$. Measurements were performed on an NV center with optical initialization and readout (green Laser pulses). The duration of the drive (~1.3 µs) was adjusted for every $\Omega$ to correspond to an integer number of $\pi$ rotation. (b) A weak signal in phase with the strong drive field drives the central transition $\omega_0$. (c) The sideband transitions can be understood as a dynamical decoupling protocol by dissecting the strong drive into a train of $\pi$ pulses and the signal into a train of weak pulses ($\epsilon$). The detuning $\Delta$ of the signal translates into periodic inversions of its axis, which are resonantly rectified by the strong drive.*

We now convert Mollow absorption spectroscopy into a pulsed protocol (Fig. 3a) to mitigate an important problem: the continuous wave (CW) protocol is prone to decoherence since fluctuations of the drive field power ($\Omega$) directly translate into frequency noise of the dressed state transition $\omega_0 \pm \Omega$. We will see that pulsed protocols shift the frequency of the Mollow sideband absorption from $\omega_0 \pm \Omega$ to $\omega_0 \pm \pi/\tau$, $\tau$ denoting the pulse spacing. Since timing ($\tau$) is controlled better than power ($\Omega$), decoherence is reduced to the intrinsic limit set by the spin qubit.

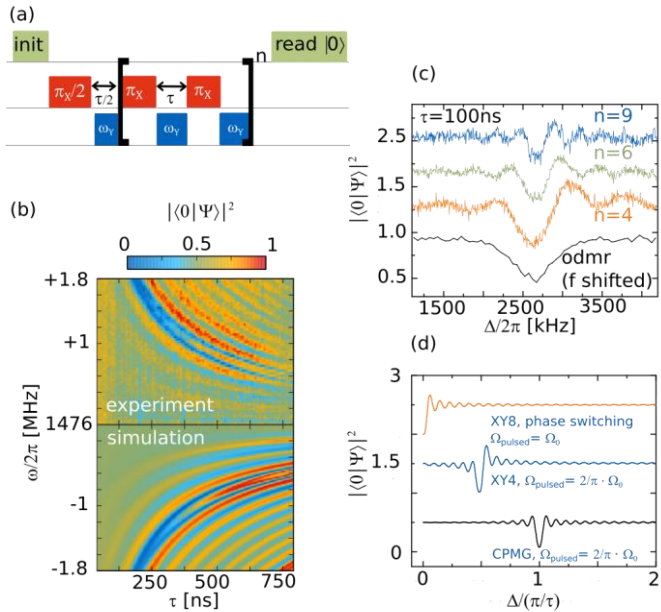

*Fig. 3 – Pulsed Mollow absorption spectroscopy. (a) Pulse sequence for high-frequency sensing, a direct implementation of the interpretation given in Fig. 2c. $\pi$ pulses at frequency $\omega_0$ emulate a strong drive to resonantly enhance a weak signal at frequency $\omega = \omega_0 + \Delta$. (b) Pulsed Mollow resonance, as measured on a NV center (upper plot) and simulated (lower plot).*

*A resonance at $\Delta = \pi/\tau$ is framed by sidebands at frequencies $\Delta = \pm\pi/\tau \pm 2\pi n/T$ (with sequence duration T, cycle number n=4). (c) Linewidth of the resonance. The line narrows below $1/T_2^*$ for sequences longer than the dephasing ($T > T_2^*$). (d) Simulated spectral response to sensing sequences with different decoupling protocols. The pulse spacing between 24 π pulses was kept constant at $\tau = 43.6$ ns. The stated effective Rabi frequencies (blue text) are for Rabi oscillations driven on resonance. 'XY8, phase switching' refers to the protocol of Fig. 4. More detailed discussion in Supplementary Note 4.*

Conversion into a pulsed protocol is best understood from tracking the spin evolution across the sideband absorption sequence (Fig 2c). We decompose the strong drive into a series of $\pi$ pulses, spaced by a time $\tau = \pi/\Omega$, and split the weak signal into a commensurate series of weak pulses with pulse area $\epsilon \ll \pi$. We equally discretize its detuning of $\Delta = \Omega$ – a continuous decrease in carrier phase – into periodic inversions of its axis, that is a discrete decrease of the phase by $\pi$ occurring with period $\pi/\Delta$. At the resonance condition $\Delta = \Omega$, this period matches the spacing $\tau = \pi/\Omega$ of the strong drive. In this case, the weak signal is resonantly rectified in the toggling frame of the spin (Fig. 2c), analogous to the situation in low-frequency sensing.

Our pulsed scheme (Fig. 3a) is an explicit implementation of this discretized sequence. We emulate the strong drive with Rabi frequency $\Omega$ by short $\pi$ pulses with a spacing $\tau = \pi/\Omega$. We find that the weak signal is most strongly absorbed at a detuning $\Delta/2\pi = \pm(2\tau)^{-1}$. The absorption resonance remains locked to this frequency as we scan $\tau$ while keeping all other parameters constant (Fig. 3b). All our observations match well with an explicit time domain simulation of the spin evolution (bottom half of Fig. 3b + Supplementary Note 3).

The bandwidth $\Delta\omega = \pi/T$ of this pulsed Mollow resonance is limited by the finite duration $T = 2n\tau$ of the sequence containing $2n$ $\pi$-pulses. Crucially, this bandwidth drops below the inhomogeneous linewidth $1/T_2^*$ if we choose a sequence longer than $T_2^*$ (Fig. 3c). The Mollow resonance is framed by sidebands at frequencies $\omega_0 \pm \pi/\tau \pm 2\pi n/T$. These are another consequence of the finite sequence length: since sensitivity is nonzero only in a rectangular window in the time domain, the sequence has a sinc response in the frequency domain. Tracing a Rabi oscillation of the weak signal along the resonance hyperbola $\Delta = \pm\pi/\tau$ we find its native Rabi frequency $\Omega_{rf}$ to be reduced to a value $\Omega_{\text{pulsed}} = 2/\pi \cdot \Omega_{rf}$. We attribute this reduction to the fact that the detuned signal, rotating at an angular frequency of $\Delta$ on the Bloch sphere, has a phase orthogonal to the strong drive phase only during a fraction of the free evolution time $\tau$. It therefore has to be scaled by a factor $(\int_0^\tau \sin(\phi(t))\,dt)/\tau$, with $\phi(t)$ denoting the phase of the signal. All of these properties are analogous to similar features in low-frequency decoupling sequences[29].

Using the time domain simulation and an analytical model (Supplementary Notes 2+4), we find Mollow resonances in many decoupling sequences, including the robust sequences CPMG, XY4 and XY8 (Fig. 3d). A detailed discussion of the decoupling sequence structure on position and shape of the resonances is given in Supplementary Note 4.

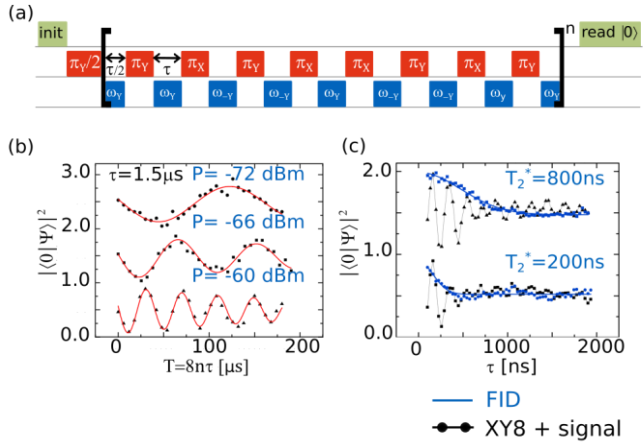

*Fig. 4 – Performance and limits of the method. (a) XY8 sequence employed to record the data of (b+c). The signal was phase-cycled in order to maximize sensitivity. (b) Rabi oscillations much slower than $T_2^*$ = 1.8 µs, where the number of π pulses was increased up to 120 while keeping the pulse spacing $\tau$ constant. (c) Signal decay for increasing pulse spacing. For spacings $\tau > T_2^*$, the pulsed Mollow protocol loses sensitivity.*

We finally demonstrate the performance and an important limit of our method by the protocol of Fig. 4a. Here, we adopt the XY8 sequence for the strong drive, in order to be maximally robust against experimental fluctuations. We phase-modulate the signal to gain a constructive contribution to the Rabi rotation during every evolution period $\tau$. In this setting we have been able to drive slow Rabi oscillations with a period as long as 100 µs (Fig. 4b). While this clearly breaks the $T_2^*$ limit in terms of signal strength, the limit reappears as a constraint on the pulse spacing $\tau$, which has to be short against $T_2^*$. For longer spacings – corresponding to slower Rabi frequencies in the CW sequence – the Mollow resonance merges with the inhomogeneously broadened transition. To verify this limit explicitly, we artificially shorten $T_2^*$ of the NV center by averaging multiple measurements taken at different, Gaussian-distributed, frequencies of the microwave drive. Tuning decoherence by this technique, we find that sensitivity breaks down if pulses are spaced by more than $T_2^*$ (Fig. 4c).

In summary, we have pushed spin-based quantum sensing to frequencies much higher than the available Rabi frequency $\Omega$. This promotes spins to phase-sensitive microwave detectors. They could operate independent of temperature and might provide sufficient sensitivity to detect single phonons and photons. Compared to competing approaches such as Josephson parametric amplifiers, our scheme has a very narrow bandwidth ($1/T_2$). This window, however, can be tuned across frequencies up to several 100 GHz by tuning the sensing spin by a magnetic field. Crucially, the absorption frequency $\omega_0 \pm \pi/\tau$ is set only by timing and frequency of the external drive, which can be controlled well. It is independent of the native spin transition and hence resilient to drifts in surrounding fields.

From a fundamental perspective, we have provided an intuitive microscopic understanding of the Mollow triplet as a pulsed quantum protocol. It appears most intriguing to extend this novel perspective to other effects of quantum interference, such as electromagnetically induced transparency (EIT).

## Methods

All experiments have been performed on single NV centers grown into a polycrystalline electronic grade IIa diamond by chemical vapor deposition (E6 part N° 145-500-0356). Both the strong drive and the weak signal were generated by an Arbitrary Waveform Generator (Rigol DG5352), which was mixed onto a GHz frequency carrier, amplified (amplifier MiniCircuits ZHL16W-43-S+), and applied to the NV center by a coplanar waveguide. The spin state was measured by fluorescence readout in a high-NA confocal microscope (excitation 532 nm, ~ 1 mW power, detection in the > 650 nm band by an objective lens Olympus UPLSAPO 60x1.35O). All sequences were recorded twice, with and without an additional $\pi$ pulse before readout. The difference of both datasets was normalized to the signal obtained in a Rabi oscillation to yield a quantitative estimate of $|\langle 0|\psi\rangle|^2$.

## Acknowledgements

This work has been supported by the Deutsche Forschungsgemeinschaft (Emmy Noether grant RE3606/1-1). The authors wish to thank Amit Finkler and Frank Deppe for helpful discussions. During redaction of this manuscript we became aware of simultaneous independent work by H. F. Fotso and V. V. Dobrovitski (arXiv:1701.07865).

## Author contributions
T.J. and F.R. conceived the experiment. A.M.W. and T.J. developed the simulations and optimized the protocols. T.J., A.M.W. and G.B. prepared the sample and performed the experiment. T.J., A.M.W. and F.R. analyzed the data. T.J., A.M.W. and F. R. wrote the paper. All authors read and commented on the manuscript.

## Competing financial interests
The authors declare not to have competing financial interests.

## Material & Correspondence

# Supplementary Information

## Supplementary Note 1: Natural qubit transition linewidth

The natural linewidth of a qubit transition is determined by the qubit dephasing time as $\Delta f_{FWHM} = 1/T_2^*$. However, the measured linewidth in an optically detected magnetic resonance (ODMR) experiment does not necessarily reflect the natural limit as several mechanisms, predominantly power broadening owing to the finite pulse bandwidth, can significantly broaden the resonance. In order to verify that the ODMR spectrum shown in Fig. 3c of the main text is not subject to parasitic broadening we test the consistency of the measured linewidth with a free induction decay (FID) experiment. As the FID is inherently unaffected by power broadening it provides a more reliable measure of the dephasing time.

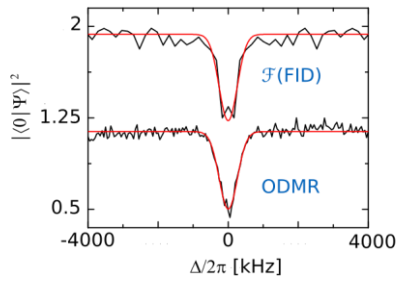

*Fig. S1* – Linewidth of the ODMR resonance. The black bottom trace is a reproduction of the ODMR curve shown in Fig. 3c of the main text. The signal amplitude was normalized to the maximum Rabi signal contrast. A Fourier transform of a time-domain FID measurement is shown in the upper trace (amplitude shifted for clarity). Red curves show fits with a Gaussian peak profile.

Fig. S1 shows the direct comparison of the Fourier transformed FID data with the measured ODMR curve. Gaussian fits to the resonance dips (red lines) yield

$$\Delta f_{FWHM,FID} = (790 \pm 50) \text{ kHz}$$

$$\Delta f_{FWHM,ODMR} = (610 \pm 14) \text{ kHz}$$

We conclude that the ODMR linewidth is in qualitative agreement with the FID curve. The linewidth of the Mollow sideband resonances shown in Fig. 3c is therefore below the natural linewidth, confirming the expectation that our sensing protocol is not limited by $T_2^*$.

## Supplementary Note 2: Description of Mollow spectra by a sensitivity function

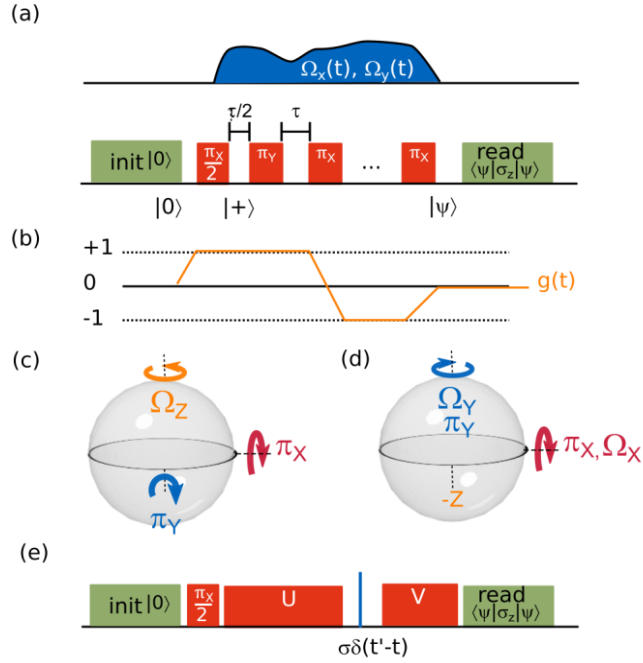

*Fig. S2 – Sensitivity function formalism. (a) We consider high-frequency sensing of a weak signal field $\Omega_x(t), \Omega_y(t)$ by a dynamical decoupling protocol involving π pulses along axes X and Y, corresponding to orthogonal carrier phases. (b) The measurement outcome $\langle\psi|\sigma_z|\psi\rangle$ can be computed as $\langle\psi|\sigma_z|\psi\rangle = \int g_{Mollow}(t)\Omega_y(t)dt$. Here g(t) is a sensitivity function modeling the change of measurement outcome induced by an infinitesimal signal applied at time t. It changes sign with every π pulse along axis X and maintains sign across π pulses along Y. (c,d) Low-frequency sensing (c) is equivalent to high-frequency sensing (d) by the identification $Z' \to Y, Y' \to -Z$. (e) Model used in the derivation. π pulses before and after an infinitesimal signal at time t are grouped to operators U and V.*

Analytically, a high-frequency sensing sequence (Fig. S2a) can be modeled by the sensitivity function formalism established for low-frequency dynamical decoupling[30]. The measurement outcome $\langle\psi|\sigma_z|\psi\rangle$ induced by a small signal field $\Omega_x(t)\sigma_x + \Omega_y(t)\sigma_y$ can be computed from projection onto a sensitivity function $g_{\text{Mollow}}(t): \mathbb{R} \to \{1, -1\}$ as $\langle\psi|\sigma_z|\psi\rangle = \int g_{\text{Mollow}}(t)\Omega_y(t)dt$ (illustrated in Fig. S2b). This is analogous to low-frequency sensing, where the rotation $\phi$ induced by a dephasing field is computed as $\phi = \int g(t)\Omega_z(t)$.

To provide a physical picture, we first review the situation of low-frequency dynamical decoupling (Fig. S2c). Here, the spin is prepared in a coherent superposition and repeatedly flipped by π pulses along axes with orthogonal carrier phases X and Y. The phase imprinted onto a superposition state $(|0\rangle + e^{i\phi}|1\rangle)/\sqrt{2}$ by a time-varying drive $\Omega_z$ is computed as $\phi = \int g(t)\Omega_z(t)$. Here, the sensitivity function g(t) changes sign with every π pulse, regardless of the pulse axis. It models the fact that $\Omega_z$ changes sign with every π pulse in the toggling frame of the spin.

High-frequency sensing (Fig. S2d) can be mapped to this situation by rotating the Bloch sphere by $\pi/2$ around axis X, identifying axes Z' with Y and Y' with –Z, respectively. Furthermore, we note that the component $\Omega_x$ of the signal induces no rotation in the limit of a very weak signal. Here, the spin never evolves far away from the axis X, remaining approximately in the eigenstates $|+\rangle, |-\rangle$ of the spin operator $\sigma_x$. In the toggling frame, the only relevant component, $\Omega_Y$, is left invariant by π pulses along Y, but changes sign with every π pulse along X. This can be modeled by a sensitivity function $g_{\text{Mollow}}(t)$ with the same properties.

The relation $\langle\psi|\sigma_z|\psi\rangle = \int g_{\text{Mollow}}(t)\Omega_y(t)dt$ can equally be derived analytically. The sensitivity of the measurement outcome $\langle\psi|\sigma_z|\psi\rangle$ to an infinitesimal signal applied at time $t$ along an axis $\boldsymbol{\sigma}$ can be formally computed as the functional derivative

$$g_{\text{Mollow}}(t) = \frac{d\langle\psi|\sigma_z|\psi\rangle}{d\,\sigma\delta(t'-t)} = \langle\psi|\sigma_z \cdot \frac{d|\psi\rangle}{d\,\sigma\delta(t'-t)} \cdot + H.C. \quad (1)$$

Grouping the control pulses before and after the infinitesimal signal to operators $\boldsymbol{U}, \boldsymbol{V}$ as illustrated in Fig. 2e, we can rewrite the derivative of the final state as

$$\frac{d|\psi\rangle}{d\,\sigma\delta(t'-t)} = -i\boldsymbol{V\sigma U}|\psi\rangle \quad (2)$$

We analyze this term separately for signals along $\boldsymbol{\sigma} = \boldsymbol{\sigma_x}$ and $\boldsymbol{\sigma} = \boldsymbol{\sigma_y}$.

The sequence is insensitive to signals along $\boldsymbol{\sigma_x}$. To see this, we note that signals are small. Therefore, the spin state $|\psi\rangle$ always remains approximately in an eigenstate of $\boldsymbol{\sigma_x}$. $|\psi\rangle = |+\rangle = (|0\rangle + |1\rangle)/\sqrt{2}$ or $|\psi\rangle = |-\rangle = (|0\rangle - |1\rangle)/\sqrt{2}$, regardless of the $\pi$ pulses applied. The action of $\boldsymbol{\sigma}$ in Eq. (2) reduces to a multiplication with $\pm 1$, which does not alter the spin state. Consequently, $\langle\psi|\sigma_z \cdot \frac{d|\psi\rangle}{d\,\sigma\delta(t'-t)} = 0$, since all matrix elements of $\boldsymbol{\sigma_z}$ involving equal spin states vanish: $\langle-|\boldsymbol{\sigma_z}|-\rangle = 0, \langle+|\boldsymbol{\sigma_z}|+\rangle=0$.

In contrast, signals along $\boldsymbol{\sigma_y}$ do alter the spin state. Their influence on the measurement outcome $\langle\psi|\sigma_z|\psi\rangle$ depends on their position in the sequence and changes sign with every $\pi$ pulse along x. To see this, we consider the cases that the infinitesimal signal is followed by a $\pi$ pulse along y or x, respectively. Due to the anticommutation relation $\{\boldsymbol{\sigma_y}, \boldsymbol{\sigma_i}\} = \delta_{yi}$, a signal along axis Y commutes with a $\pi$ pulse along Y. It anticommutes with a $\pi$ pulse along axis X so that it has opposite effects before and after a $\pi_x$ pulse. Accordingly, the sensitivity function $g_{\text{Mollow}}(t)$ has to maintain its value across a $\pi$ pulse along Y, and has to change sign with every $\pi$ pulse along X.

## Supplementary Note 3: Time domain simulations

All simulations are based on the semi-classical Rabi Hamiltonian

$$\hat{H} = \frac{\hbar}{2}(\Delta \sigma_z + \Omega \left(\cos(\phi)\, \boldsymbol{\sigma_x} + \sin(\phi)\, \boldsymbol{\sigma_y}\right))$$

where $\Delta = \omega_0 - \omega$ is the detuning of the resonant qubit two-level transition $\omega_0$ from the coupled electromagnetic field with frequency $\omega$, phase $\phi$ and resonant Rabi frequency $\Omega$.

For every pulse in the sequence we calculate the time evolution operator, which is determined by the unified Rabi frequency $\hat{\Omega}_u$.

$$\hat{U} = \exp(-\frac{i}{2}\hat{\Omega}_u\, t)$$

If only one electromagnetic field couples to the two-level system, the well-known expression for the unified Rabi frequency in rotating wave approximation is

$$\hat{\Omega}_u^* = \Omega\cos(\phi)\, \boldsymbol{\sigma_x} + \Omega\sin(\phi)\, \boldsymbol{\sigma_y} + \Delta \boldsymbol{\sigma_z}$$

However, in our setting two fields must be considered. The weak signal field $\omega_{rf}$ is detuned from the spin transition by $\Delta_{rf} = \omega_0 - \omega_{rf}$. In addition, we have a resonant dressing field which correspondingly is detuned from the weak field by $\Delta_{rf,dd} = \omega_{rf} - \omega_{dd}$ (where generally $\omega_{dd} \approx \omega_0$ and hence $\Delta_{rf,dd} \approx -\Delta_{rf}$).

For practical purposes, we now switch to the rotating frame of the weak signal field where the respective expressions are

$$\hat{\Omega}_{rf} = \Omega_{rf} \cos(\phi_{rf})\, \boldsymbol{\sigma_x} + \Omega_{rf} \sin(\phi_{rf})\, \boldsymbol{\sigma_y} + \Delta_{rf} \boldsymbol{\sigma_z}$$

$$\hat{\Omega}_{dd} = \Omega_{dd}\cos(\phi_{dd} + \Delta_{rf,dd} t)\, \boldsymbol{\sigma_x} + \Omega_{dd}\sin(\phi_{rf} + \Delta_{rf,dd} t)\, \boldsymbol{\sigma_y} + \Delta_{rf} \boldsymbol{\sigma_z}$$

We note that in this frame the rotation axis of the drive field precesses at a speed determined by the detuning $\Delta_{rf,dd}$. Assuming sharp control pulses, we approximate the expression for the drive field using only the starting time $t_i$ of each sharp pulse

$$\hat{\Omega}_{dd} \approx \Omega_{dd}\cos(\phi_{dd} + \Delta_{rf,dd} t_i)\, \boldsymbol{\sigma_x} + \Omega_{dd}\sin(\phi_{rf} + \Delta_{rf,dd} t_i)\, \boldsymbol{\sigma_y} + \Delta_{rf} \boldsymbol{\sigma_z}$$

In this way, we gain two unified Rabi frequencies $\hat{\Omega}_{rf}, \hat{\Omega}_{dd}$ which we apply for the respective pulses.

$$\hat{\Omega}_u = \begin{cases} \hat{\Omega}_{rf} & \text{for signal pulses} \\ \hat{\Omega}_{dd} & \text{for drive pulses} \end{cases}$$

The total time evolution of the spin with initial state $|0\rangle$ under a sequence of pulses $\{U_0(t_0), \ldots, U_n(t_n)\}$, $t_n > t_0$  is then given by

$$|\Psi_f\rangle = \prod_{i=n}^{0} \hat{U}_i\, |0\rangle$$

## Supplementary Note 4: Robust Pulse Sequences

As shown in the main text, long sequences are favorable for best sensitivity. Since $T_2^*$ sets an upper limit to the pulse spacing $\tau$, this requires application of as many $\pi$ pulses in the strong dressing sequence as possible. Owing to their capability to correct for pulse errors and environmental fluctuations, dynamical decoupling (DD) protocols are an established technique for implementing such long control pulse trains.

In principle, all DD protocols that contain $\pi$ pulses with a carrier phase orthogonal to the phase of the signal are compatible with our RF sensing scheme (see discussion of the sensitivity function in Supplementary Note 2). However, if phase control of the signal field is not assumed, an additional constraint on the periodicity of the control pulses arises: Since phase switching is effectively obtained by a constant detuning $\Delta$, such orthogonal pulses (e.g. $\pi_X$ for a signal phase Y) must be equally spaced in the control sequence. Consequently, XY8 cannot be applied under this assumption.

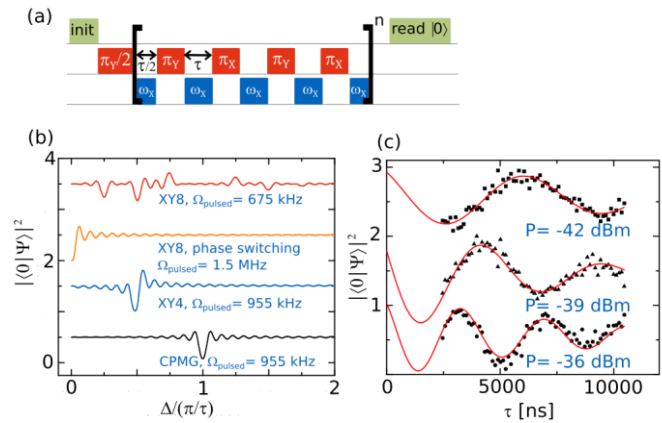

*Fig. S3 – Influence of dynamical decoupling dressing sequence. (a) XY4 sequence employed to record the data shown in (c). (b) Simulation of the spectral response to sensing sequences with different decoupling protocols (reproduced from Fig. 3d with added data). All simulations use 24 $\pi$ pulses with a fixed pulse spacing $\tau$=43.6ns. The signal Rabi frequency was set to $\Omega_{rf} = 1.5 * 10^6$ Hz, yielding effective Rabi frequencies $\Omega_{XY8,ps} = \Omega_{rf}$, $\Omega_{XY8} = \frac{\sqrt{2}}{\pi}\Omega_{rf} \approx 0.45\Omega_{rf}$, $\Omega_{XY4} = \Omega_{CPMG} = \frac{2}{\pi}\Omega_{rf} \approx 0.64\Omega_{rf}$. (c) Rabi oscillations measured along the hyperbola defined by $\Delta = \frac{\pi}{2\tau}$.*

As shown in Fig. S3b and Fig. 3d of the main text, a different spectral response arises depending on the distinct DD protocol used: XY8 without phase switching of the weak signal (red curve) shows a complex sideband structure with multiple strongly damped Mollow resonances, reflecting the different periodicities in the sequence. On the most pronounced resonance at $\Delta = \pi/2\tau$, a reduced effective Rabi frequency of $\Omega_{XY8} = \frac{\sqrt{2}}{\pi}\Omega_{rf}$ is observable. However, by adding a suitable phase switching (orange curve, see Fig. 4a for the pulse protocol), we can make the central Mollow peak sensitive to the weak field, i.e. drive weak Rabi oscillations with frequency $\Omega_{rf}$ at $\Delta = 0$.

In the case of a CPMG dressing sequence (protocol shown in Fig. 3a), $\pi$ pulses with a phase orthogonal to the signal field are equidistantly spaced. Consequently, weak Rabi oscillations with effective frequency $\Omega_{CPMG} = \frac{2}{\pi}\Omega_{rf}$ can be driven in the Mollow resonance at detuning $\Delta = \pi/\tau$ (black curve in Fig. S3b).

The effect of the spacing between orthogonal pulses becomes clear when we compare this to the response to a XY4 protocol as shown in Fig. S3a. Here, only the phase of every second $\pi$ pulse is orthogonal to the signal phase. As a result, we observe (blue curve in Fig. S3b) that the Mollow resonance is shifted to $\Delta = \pi/2\tau$, that is half of the detuning at which the CPMG resonance appears. By contrast, the effective Rabi frequency $\Omega_{XY4} = \frac{2}{\pi}\Omega_{rf}$ is identical.

To confirm the behavior observed in our simulations, we also obtained experimental data for XY4, where we controlled the detuning such that Rabi oscillations along the hyperbola $\Delta = \pi/2\tau$ could be measured (Fig. S3c). The data shows excellent agreement with the predicted evolution, confirming the applicability of our model.